\documentclass[twocolumn,showpacs,preprintnumbers,prl,aps,amssymb,superscriptaddress]{revtex4} 
\usepackage{graphicx} 
\usepackage{dcolumn} 
\usepackage{bm} 
\begin{document} 
\newcommand{\Ir}{CeIrIn$_5$} 
\newcommand{\ie}{{\it i.e.}} 
\newcommand{\eg}{{\it e.g.}} 
\newcommand{\etal}{{\it et al.}} 
  
\title{Heat transport study of Field-Tuned Quantum Criticality in CeIrIn$_5$} 
  
\author{H.~Shakeripour}  
\email{hshakeri@cc.iut.ac.ir}  
\affiliation{Department of Physics, Isfahan University of Technology, Isfahan 84156-83111, Iran} 
\affiliation{D\'epartement de physique and RQMP, Universit\'e de Sherbrooke, Sherbrooke, Canada} 

\author{M.~A.~Tanatar} 
\affiliation{D\'epartement de physique and RQMP, Universit\'e de Sherbrooke, Sherbrooke, Canada} 

\author{C.~Petrovic} 
\affiliation{Department of Physics, Brookhaven National Laboratory, Upton, New York 11973, USA} 
\affiliation{Canadian Institute for Advanced Research, Toronto, Ontario, Canada} 
\author{Louis Taillefer} 
\email{Louis.Taillefer@USherbrooke.ca} 
\affiliation{D\'epartement de physique and RQMP, Universit\'e de Sherbrooke, Sherbrooke, Canada} 
\affiliation{Canadian Institute for Advanced Research, Toronto, Ontario, Canada} 
\date{\today} 
  
\begin{abstract} 
The in-plane electrical resistivity, $\rho$, and thermal conductivity, $\kappa$, of heavy fermion superconductor CeIrIn$_5$ were measured down to 40 mK in the magnetic fields up to 11 T applied along the $c$-axis. For all fields above $H_{c2}$=4~T of filamentary superconductivity, we find that the ratio of heat and charge conductivities in $T \to$~0 limit obeys the Wiedemann-Franz law, $\kappa/T=L_0/ \rho$, where $L_0$=2.45$\times$10$^{-8}$ $W \Omega K^{-2}$ is Sommerfeld value of Lorenz number. The temperature-dependent parts of both the electrical resistivity and of the thermal resistivity, $w\equiv T/L_0 \kappa$, follow functional dependence expected for Fermi liquid theory of metals with $\rho-\rho_0=AT^2$, $w -w_0=B T^2$, with $\rho_0=w_0$ and $B \approx 2A$. The coefficient $B$ does not show significant field dependence even on approaching $H_{c2}$=0.4~T of bulk superconducting state. Weak response to magnetic field is in stark contrast with the behavior found in a closely related CeCoIn$_5$, in which field-tuned quantum critical point coincides with $H_{c2}$. The value of the electron-electron mass enhancement, as judged by $A$ and $B$ coefficients, is about one order of magnitude reduced in CeIrIn$_5$ as compared to CeCoIn$_5$ (in spite of two times larger zero-field $\gamma_0$ in CeIrIn$_5$ than $\gamma_0$ in CeCoIn$_5$), suggesting that the material is significantly further away from magnetic quantum critical point at bulk $H_{c2}$ and at all fields studied. Suppressed Kadowaki-Woods ratio in CeIrIn$_5$ compared to CeCoIn$_5$ suggests notably more localized nature of $f$-electrons in the compound. 
\end{abstract} 
\maketitle

\section{Introduction} 
Ce based heavy fermion compounds CeMIn$_5$ (M=Co, Rh, Ir), known as the 115 family, show a variety of ground states at ambient pressure. CeIrIn$_5$ and CeCoIn$_5$ are bulk superconductors with the transition temperatures $T_c$ = 0.4~K and 2.3~K, respectively \cite{Petrovic-Ir,Petrovic-Co}. Filamentary superconductivity is observed in CeIrIn$_5$ below $T_{cf} \sim$ 1.2~K \cite{Petrovic-Ir}. CeRhIn$_5$, an antiferromagnet with Neel temperature $T_N$ = 3.8~K, shows ambient pressure superconductivity at very low temperatures, $T_c \approx$ 0.1~K \cite{Deguchi,PaglioneRhSC} and with the same value of $T_c \approx$ 2.5~K as CeCoIn$_5$ under pressure of 2 GPa \cite{Hegger}.

Magnetic order can be induced in the superconducting members by Cd and Hg substitutions of In \cite{Cddoping}, so that phase diagrams with rich interplay of magnetism, quantum critical points (QCP) at $T_N \to 0$ \cite{Mathur} and superconductivity are found as functions of pressure and composition (see Refs. \onlinecite{Tompson,Sarrao} for reviews). Doping studies found that CeIrIn$_5$ \cite{Pagliuso,Ronning2014,qcp}, similar to CeCoIn$_5$ \cite{Sidorov,Cddoping}, also lies in the vicinity of the magnetic quantum critical point, and thus its superconductivity is most likely magnetically mediated. Our studies of thermal conductivity in both CeCoIn$_5$ \cite{unpaired} and CeIrIn$_5$ \cite{hybrid,universal}, though, found significant deviations from simple $d$-wave scenario, supported by many experiments \cite{Izawadwave115,Sakakibaradwave,Davisdwave,Yazdanidwave}.

Magnetic field can also be used as the non-thermal tuning parameter for quantum critical point, and in CeCoIn$_5$ field-tuned QCP is mysteriously coinciding with the superconducting upper critical field for orientations of magnetic field along both $c$-axis \cite{Paglione-QCP,Bianchi} and $ab$-plane \cite{Ronning} of the tetragonal crystals. The existence of QCP is reflected in anomalous temperature dependences of electrical resistivity \cite{Petrovic-Co,Sidorov,WFScience} closely following $T$-linear dependence, of the specific heat showing logarithmic divergence, in band-dependent divergence of cyclotron effective masses \cite{Shishido,Julian} and directional violation of Wiedemann-Franz law \cite{nonvanishing,WFScience}.

In CeRhIn$_5$ field-induced magnetic order is found in the pressure range below $p_{max}$=2.5 GPa, corresponding to maximum superconducting $T_c$ \cite{Knebel}. Interestingly, similar phenomenology of QCP coinciding with the first order transition at $H_{c2}$ is found in closely related heavy fermion Ce$_2$PdIn$_8$ \cite{ShiyanCePdIn}, as well as in some other unconventional superconductors, including cuprates \cite{PaglioneButch} and iron-based material KFe$_2$As$_2$ \cite{QCPKFe2As2}.

Of great interest is the question if field tuned quantum criticality can be found in CeIrIn$_5$. The two materials, CeCoIn$_5$ and CeIrIn$_5$, have identical Fermi surfaces \cite{Settai,bandstructureFS} and show very similar normal state properties \cite{Settai}, so it is natural to expect similarity in response to magnetic field. Despite this apparent similarity, heat capacity measurements \cite{Stewart,Capanheatcapacity,Kittaka} do not indicate critical behavior near bulk upper critical field, finding nearly temperature-independent $C/T$ at the lowest temperatures. Instead, heat capacity, torque and resistivity studies suggest that CeIrIn$_5$ shows metamagnetic transition with critical field of 25~T, far above superconducting $H_{c2}$ \cite{Capanheatcapacity,Capanmagnetization,Kim}. These studies, though, reveal several unusual features. While the electronic term $\gamma_0 \equiv C/T$ in heat capacity saturates at constant value below $T_{FL}\approx$ 0.8~K in magnetic field of  1~T and remains nearly constant with magnetic field $H \parallel c$ in 1~T to 17~T range, with $\gamma_0 \approx$ 0.7 to 0.8 J/mole K$^2$ \cite{Capanheatcapacity,Kittaka}, deviations from this value at high fields start at progressively lower $T_{FL}$ \cite{Capanheatcapacity}. Another study suggested that the superconducting dome of CeIrIn$_5$ is located inside the dome of precursor phase, similar to pseudogap state in the high $T_c$ cuprates \cite{precursor,precursor2}, suppressed at a field of 6~T.

We should keep in mind though that resistivity studies in CeIrIn$_5$ cannot probe the region close to bulk $H_{c2}$, hidden inside the domain of filamentary superconductivity \cite{Petrovic-Ir,Bianchi2001}. Subtraction of Schottky anomaly in heat capacity measurements always leaves some uncertainty in the behavior at the lowest temperatures. To get an additional insight into response of CeIrIn$_5$ to application of magnetic field, here we report field evolution of the electronic transport properties of CeIrIn$_5$ using simultaneous electrical resistivity and thermal conductivity measurements, with the latter allowing to probe the range of filamentary superconductivity close to bulk $H_{c2}$. We find that the Wiedemann-Franz law is universally obeyed in $T \to $~0 limit for all fields where resistivity measurements are not affected by filamentary superconductivity, proving true bulk nature of both electrical and heat transport. For all fields above bulk $H_{c2}$ we observe a Fermi-liquid temperature dependence of thermal resistivity, $w-w_0=BT^2$, slightly decreasing with field $B$ coefficient and nearly field-independent $T_{FL}$. Notably smaller values of $B$ in CeIrIn$_5$ when compared to CeCoIn$_5$, as well as their very mild dependence on magnetic field, suggest that CeIrIn$_5$ is significantly further away from magnetic instability at $H_{c2}$ and the field tuned criticality is significantly different in the two materials. We link this difference to anomalously low value of Kadowaki-Woods ratio of the $T^2$ coefficient in electrical resistivity versus $\gamma$ coefficient of the specific heat. This suggests more localized character of magnetism in CeIrIn$_5$ than in CeCoIn$_5$.

\section{Experimental} 

Single crystals of CeIrIn$_5$ were grown using self-flux method \cite{Petrovic-Ir}. As-grown crystals were of thin plate-like shape, with large surfaces corresponding to the (001) basal plane of the tetragonal lattice. Two samples used in this study were from two different batches, but revealed nearly identical results. Samples were cut into rectangular shape with dimensions of $\sim 4 \times 0.1 \times 0.045$~mm$^3$ and $\sim 2 \times 0.1 \times 0.01$~mm$^3$. Four contacts to the samples were made by soldering 50 $\mu m$ silver wires with indium solder, which resulted in the contact resistance typically $\sim 1$~m$\Omega$ at low temperature. Same contacts were used to measure electrical resistivity and thermal conductivity in a dilution refrigerator, thus essentially eliminating uncertainty of the geometric factor determination. The thermal conductivity was measured using a standard four-probe steady-state method with two RuO$_2$ chip thermometers calibrated {\it in situ} against a reference Ge thermometer. The electrical resistivity was measured using LR700 ac resistance bridge, operating at a frequency of 16 Hz, by applying 0.1 mA excitation currents. To enable low-noise measurements of low resistance samples in identical conditions with thermal conductivity studies, measurements were performed in a separate dilution refrigerator run, during which the high-resistance wire coils of the thermal conductivity apparatus were temporary shorted with silver wires, see Ref.~\onlinecite{WFScience} for details.

The phonon contribution to thermal conductivity, $\kappa_g$, was determined from electrical resistivity and thermal conductivity measurements in heavily disordered samples of Ce$_{1-x}$La$_x$IrIn$_5$ with $x$=0.2 and residual resistivity $\sim 10~\mu \Omega$cm, see Ref.~\onlinecite{Paglione-Rh} for details. In these heavily disordered samples temperature-dependent inelastic scattering contribution to resistivity can be neglected, and electronic contribution can be directly determined from Wiedemann-Franz law as $\kappa_e/T=L_0/\rho$, which gives $\kappa_g=\kappa-\kappa_e$. 
This contribution represents minor correction in the temperature range below 1~K, the focus of this study.

\section{Results}


\begin{figure*} 
\centerline{ 
\scalebox{0.47}{ 
\includegraphics{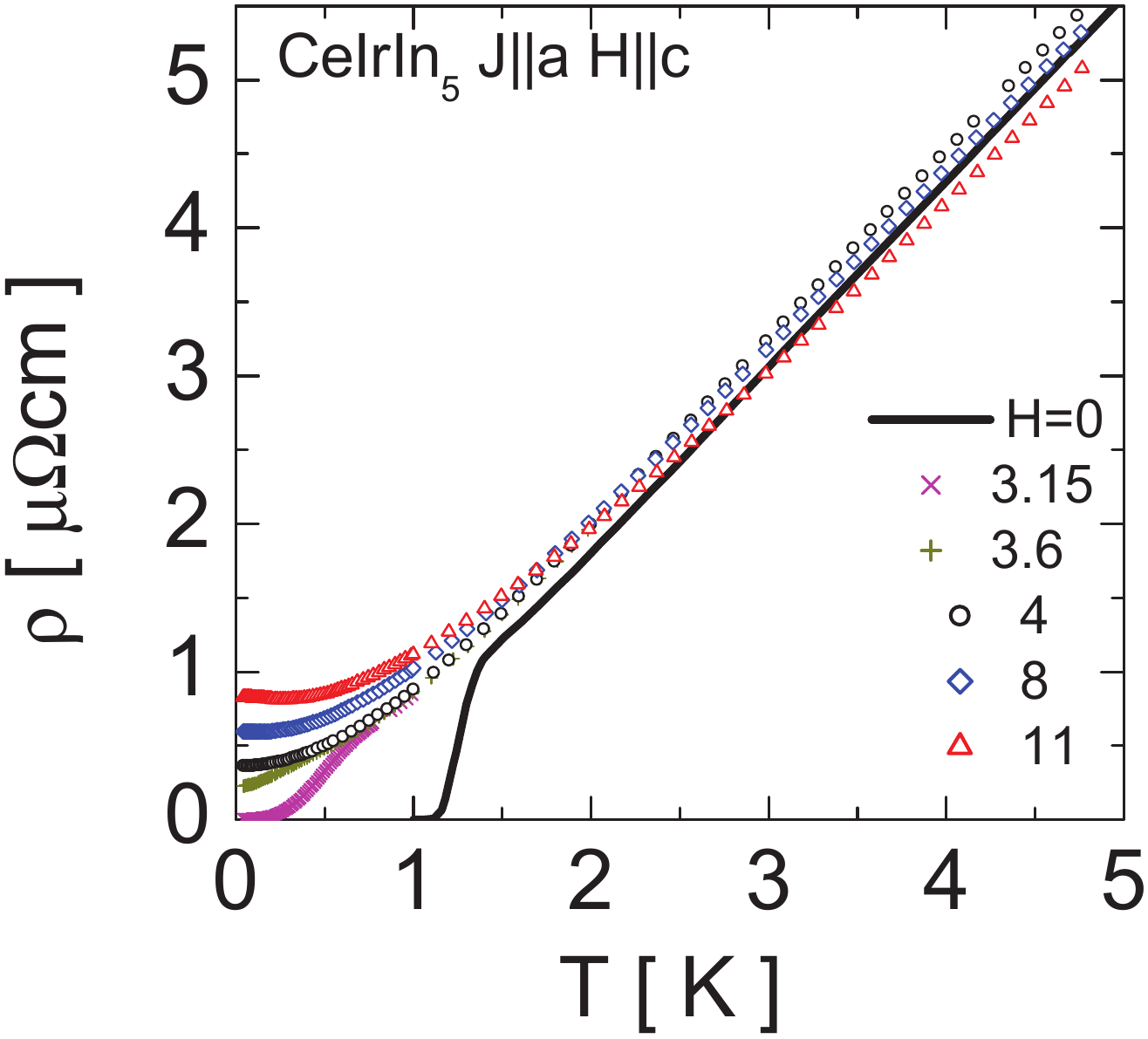}} 
\scalebox{0.47}{ 
\includegraphics{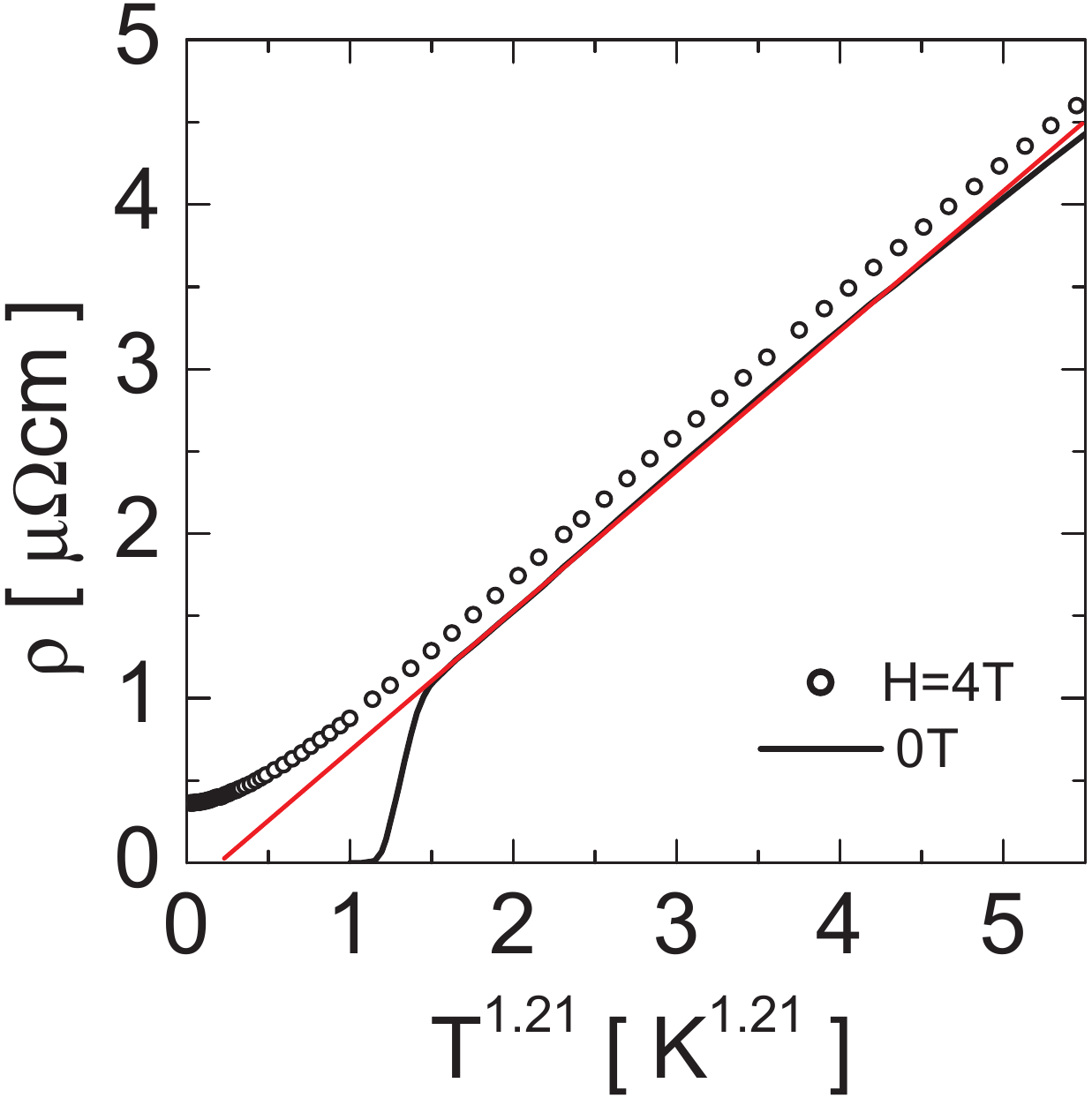}}} 
\centerline{ 
\scalebox{0.47}{ 
\includegraphics{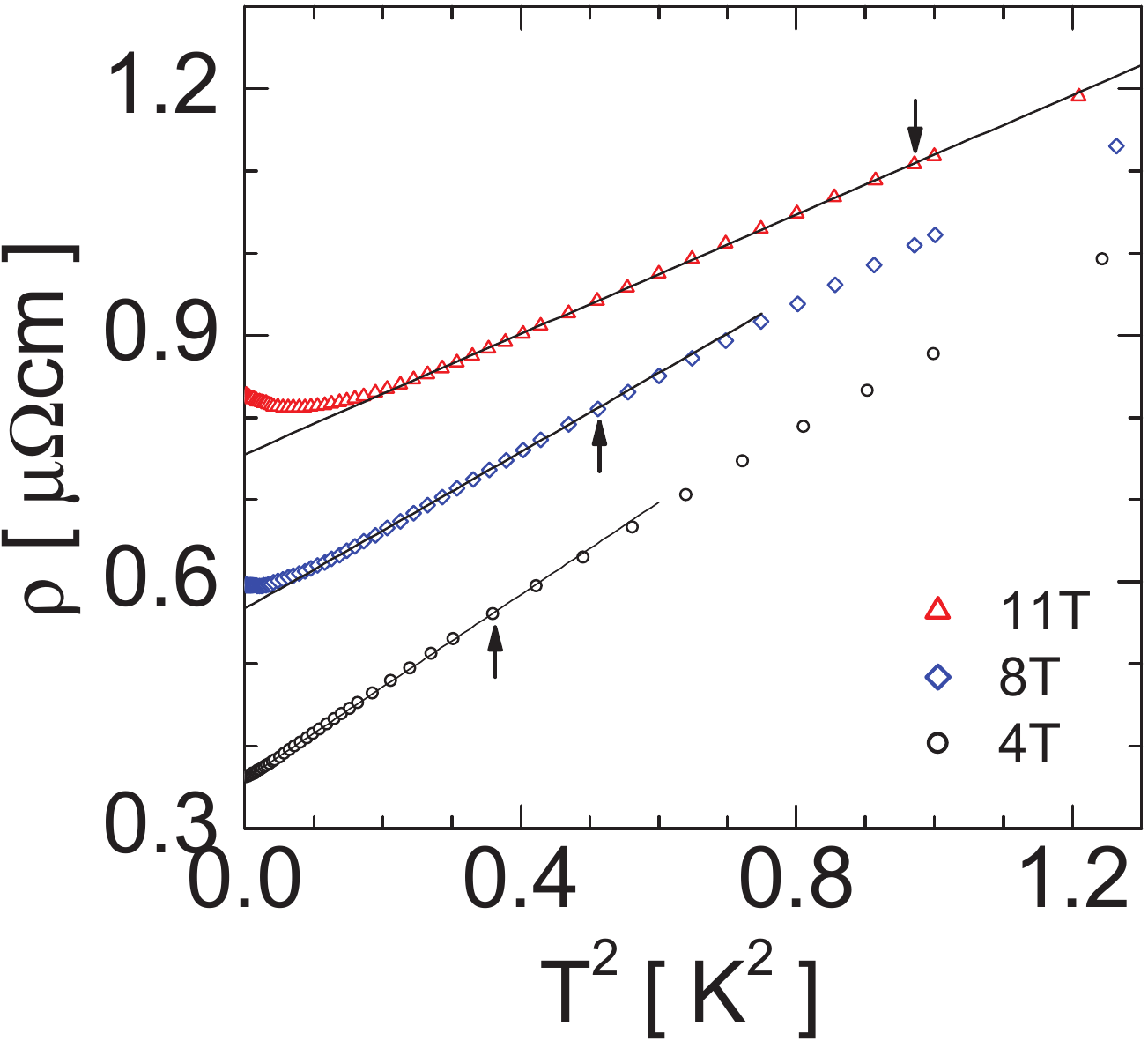}} 
\hspace*{0.21cm} 
\scalebox{0.47}{ 
\includegraphics{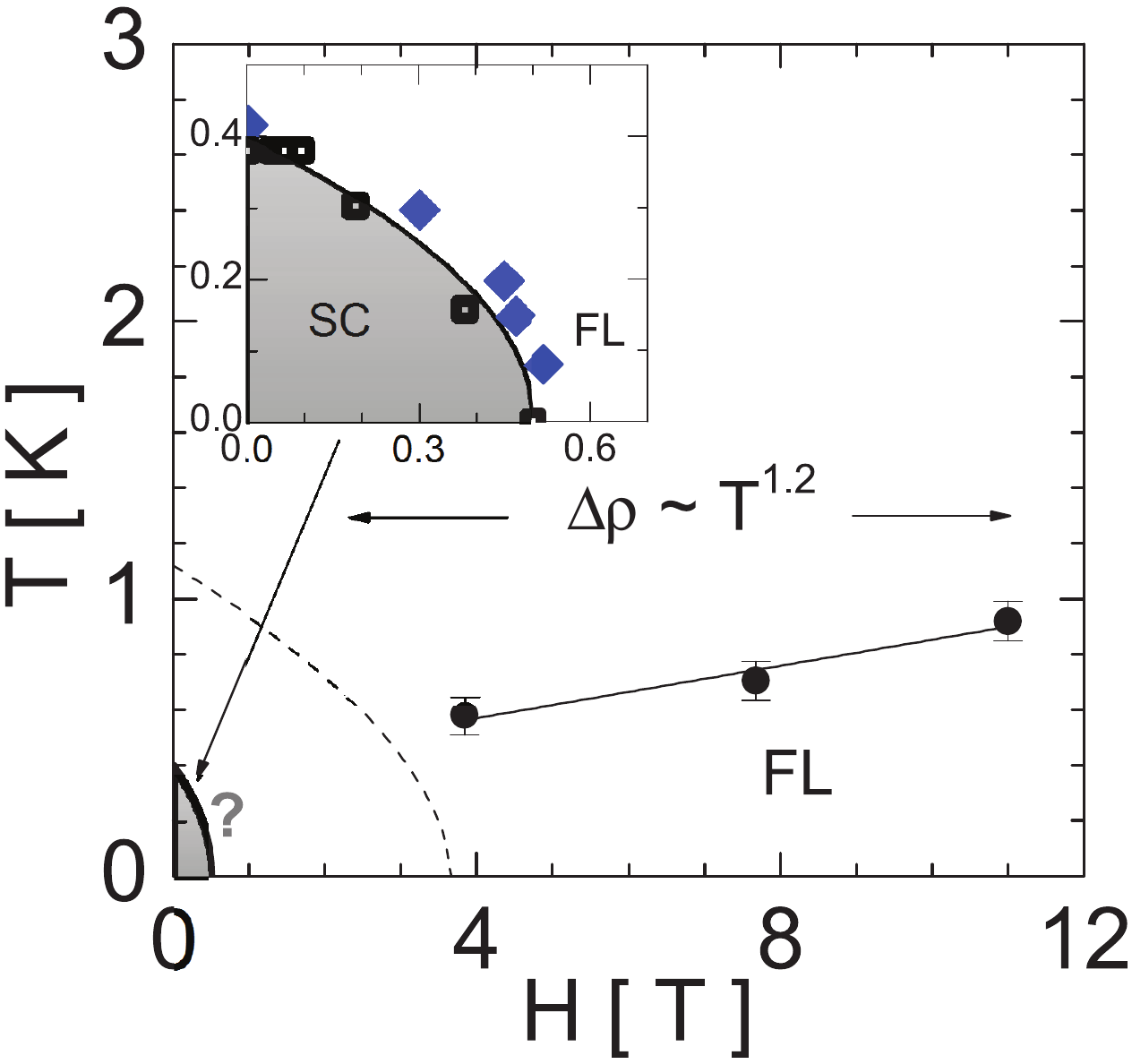}}} 
\caption{\label{fig:resistivity}  
Temperature-dependent in-plane electrical resistivity of CeIrIn$_5$ measured in magnetic fields of 0, 3.15, 3.6, 4, 8, 11 T applied along the tetragonal $c$-axis (top left panel). The filamentary superconductivity, observed in zero field below $T_{cf}=$1.2~K, is completely suppressed by $H_{c2f}\approx$ 4~T, revealing saturating $\rho(T)$ dependence on $T \to $~0. Top right panel shows the data in zero field and in field of 4~T plotted vs. $T^n$ with $n$=1.21, best fit to the power-law function above the filamentary superconducting transition at $T_{cf}$=1.2~K. Bottom left panel shows $\rho (T)$ data plotted vs. $T^2$ for magnetic field $H>H_{c2f}$. Linear dependence, found at low temperatures for all fields, suggests the validity of the Fermi-liquid picture predicting $\rho=\rho_0+AT^2$. Arrows indicate the high temperature end of the interval where this linear dependence is obeyed, marking a characteristic temperature $T_{FL}$. Note low values of $T_{FL}\approx$ 0.7 to 1~K for all fields and the slowly decreasing slope of the $\rho (T^2)$ curves with magnetic field, suggesting decreasing $A$ coefficient (and thus effective mass) with increasing magnetic field. Bottom right panel summarizes phase diagram as found from resistivity, thermal conductivity (black squares) \cite{hybrid} and heat capacity (blue diamonds) \cite{Kittaka} measurements. Dashed line in the main panel denotes domain of filamentary superconductivity. We also show field-evolution of the Fermi-liquid properties as found from resistivity measurements (black dots) above $H_{c2f}$. 
} 
\end{figure*}  

\begin{figure*} 
\centerline{ 
\scalebox{0.47}{ 
\includegraphics{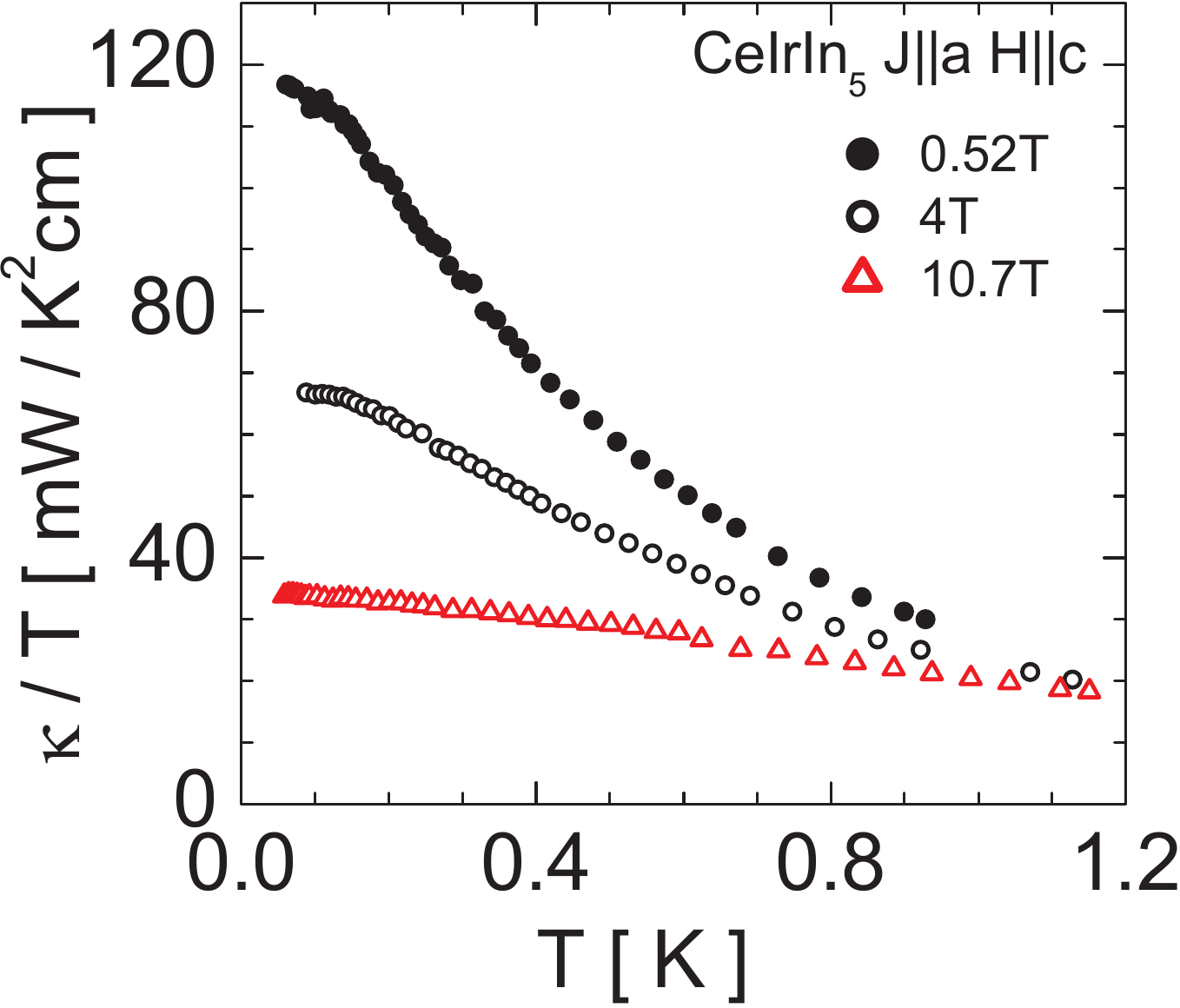}} 
\scalebox{0.47}{ 
\includegraphics{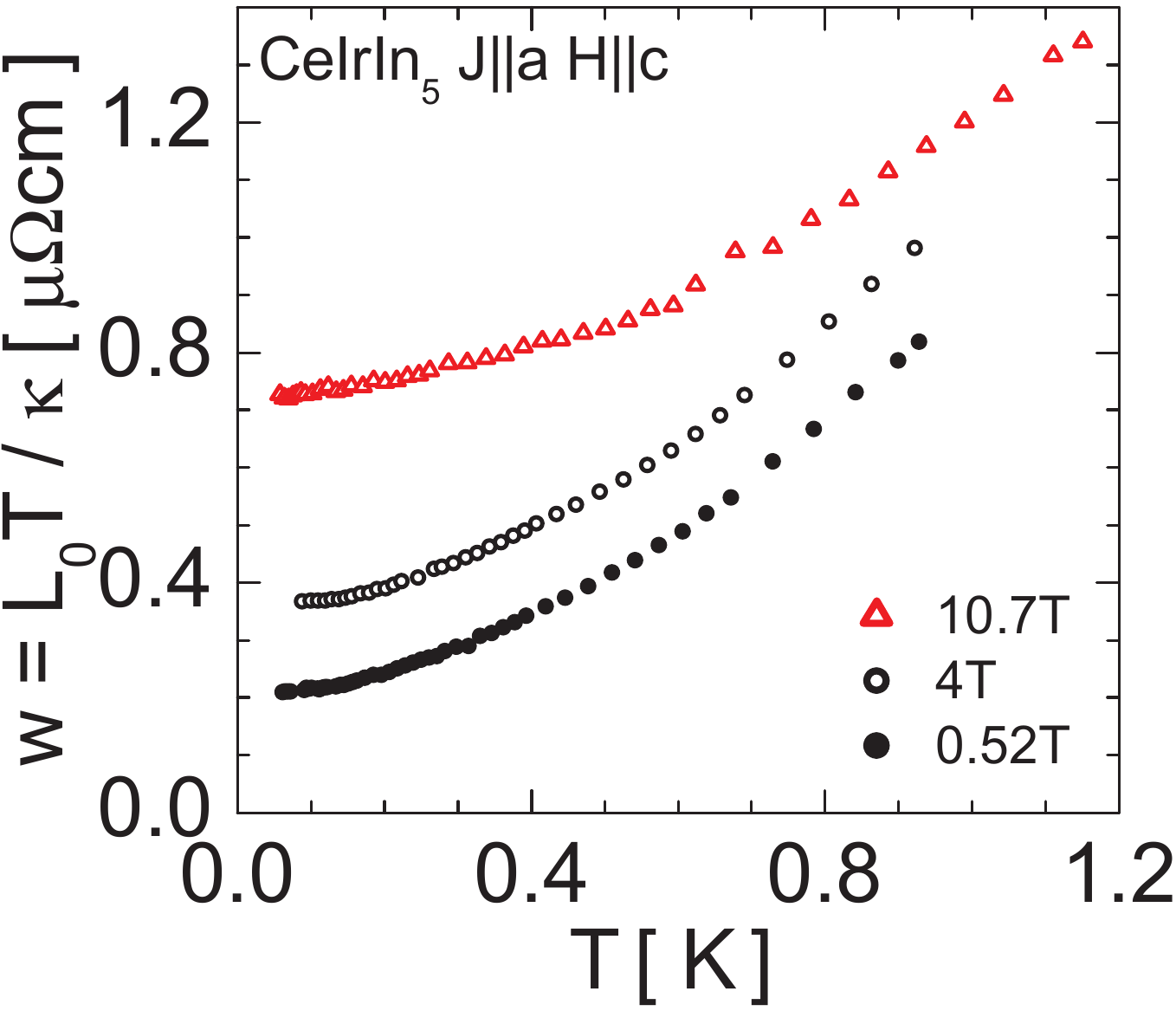}}} 
\centerline{ 
\scalebox{0.47}{ 
\includegraphics{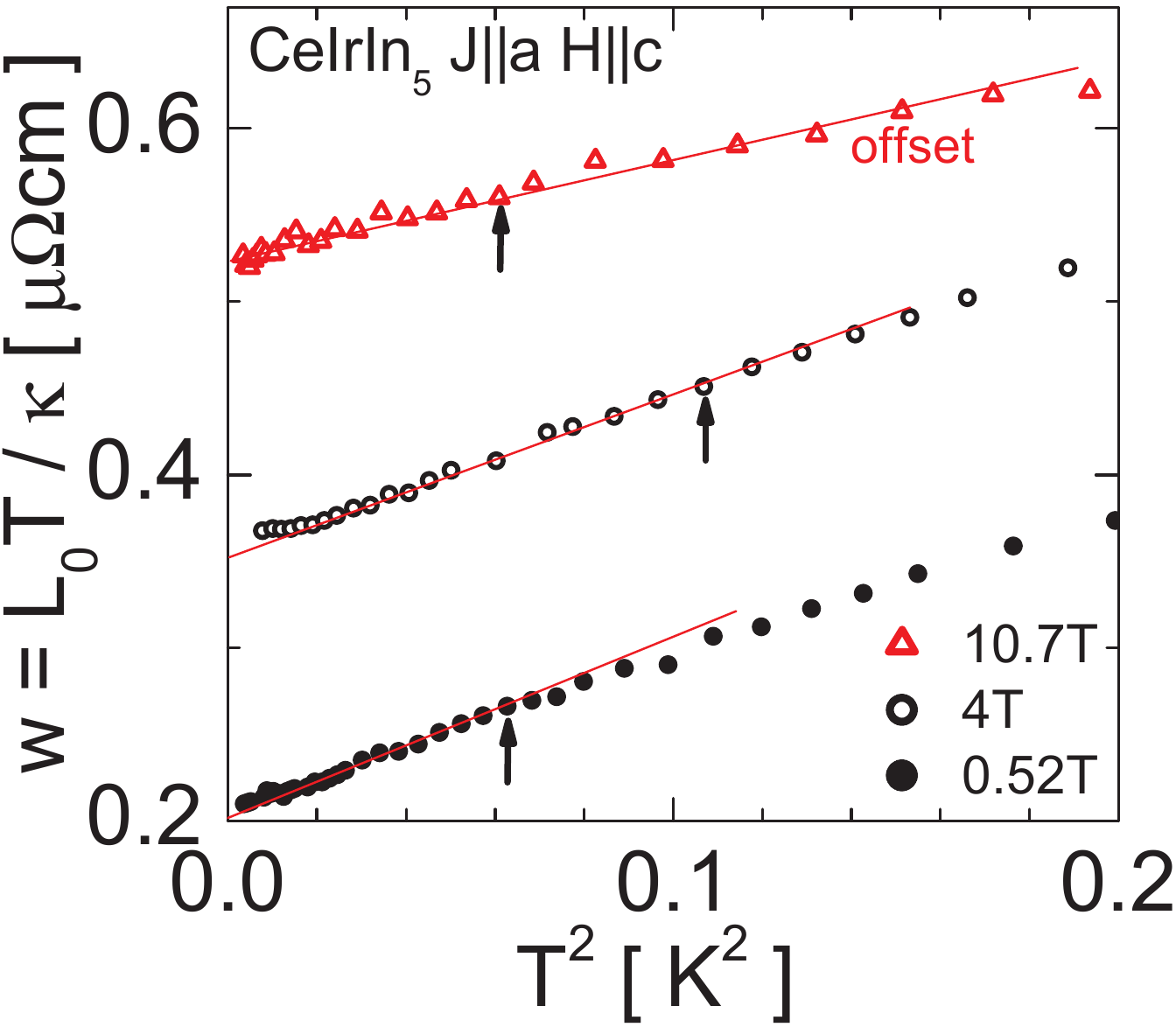}} 
\scalebox{0.47}{ 
\includegraphics{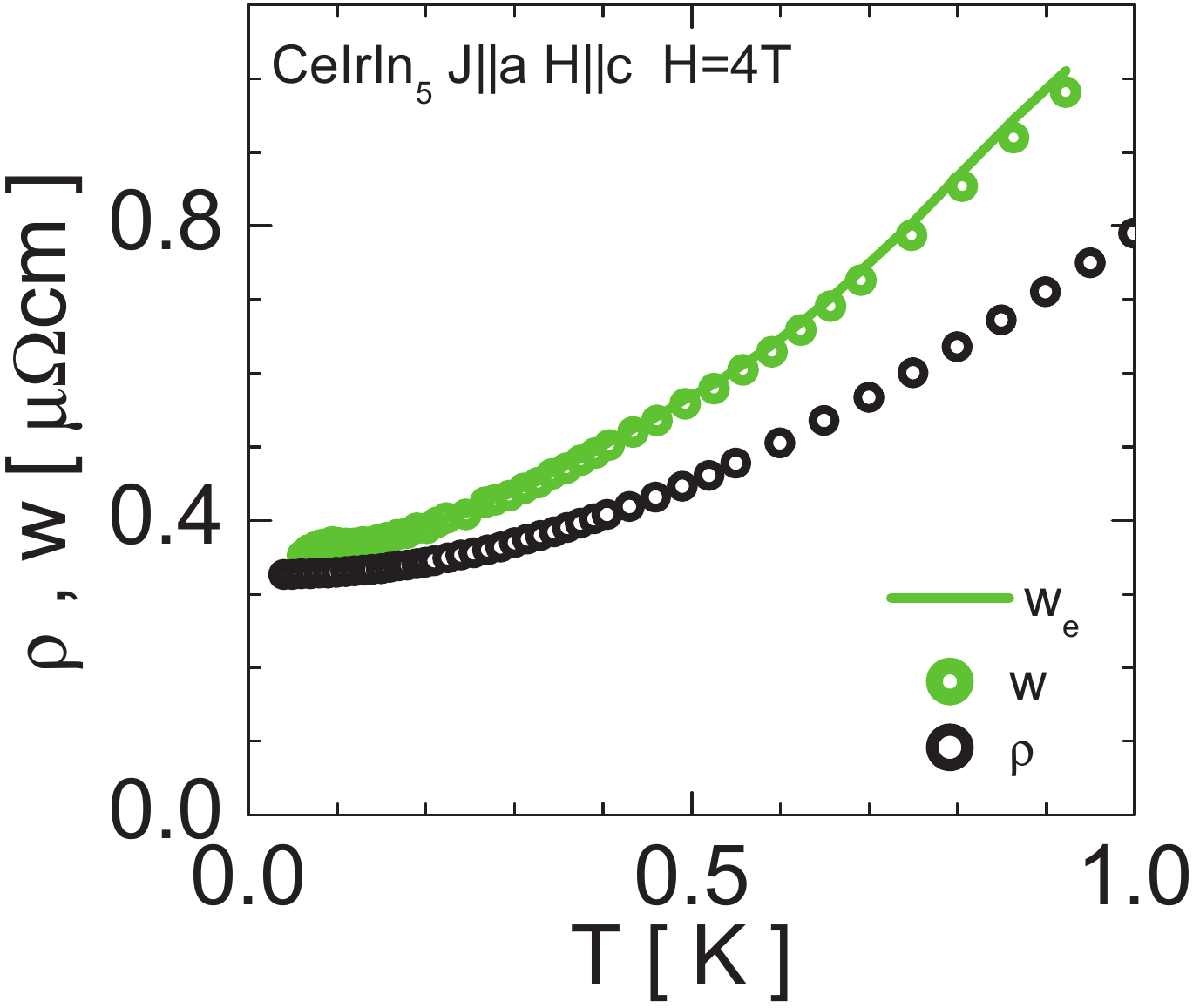}}} 
\caption{\label{fig:heat} 
Temperature-dependent in-plane thermal conductivity of CeIrIn$_5$, plotted as $\kappa/T$ vs. $T$, measured in magnetic fields of 0.52~T (slightly above bulk $H_{c2}=$0.49~T), 4~T and 10.7~T applied along the tetragonal $c$-axis (top left panel). The filamentary superconductivity does not affect thermal conductivity of the samples, but leads to extrinsic violation of the Wiedemann-Franz law. On complete suppression of filamentary superconductivity in magnetic field 4~T, the thermal resistivity $w\equiv L_0T/\kappa$, closely matches electrical resistivity $\rho$ in $T=$ 0 limit, as shown in top right panel, clearly satisfying Wiedemann-Franz law. Moreover, in the temperature range $T<1~K$, in which phonon contribution is negligible, thermal resistivity $w_e$ follows the same Fermi-liquid temperature dependence $w=w_0+BT^2$. This dependence is closely obeyed for all magnetic fields studied (bottom panels), even in very vicinity of bulk $H_{c2}$. Due to larger noise in the data, determination of $T_{FL}$ in $w(T^2)$ plots is even less precise than from the resistivity data (bottom right panel). } 
\end{figure*}

In Fig.~\ref{fig:resistivity} we plot temperature-dependent resistivity of CeIrIn$_5$ measured in magnetic fields of 0, 3.15, 3.6, 4, 8, 11 T applied along the tetragonal $c$-axis (top left panel). In zero field, the filamentary superconductivity is observed below $T_{cf}=$ 1.2~K, the resistive transition is completely suppressed in field of 4~T. The filamentary superconductivity was shown to have the same angular dependence of $H_{c2}$ as the bulk one \cite{Bianchi2001,Settai} and suggested to originate from strain in the samples, likely induced by inclusions of different phase \cite{ThompsonSteglich}. Similar phenomenon is known in Sr$_2$RuO$_4$ in which superconductivity with higher $T_c$ is induced by  inclusions of Ru-metal \cite{Sr2RuO4Hc2} and in some other materials including chain Nb$_2$Se$_3$ \cite{Nb2Se3}. 

The $\rho(T)$ dependences at temperatures above 1~K are very similar for all magnetic fields. They clearly deviate from the functional form expected in Fermi-liquid theory, $\rho=\rho_0+AT^2$. The residual electrical resistivity, $\rho_{0}$, of the samples was determined from measurements down to 40 mK in the magnetic field of 4~T, which was sufficient to completely suppress filamentary superconductivity. It was about $\sim 0.3~\mu\Omega$~cm, showing high purity of the samples. This value is notably lower than $\sim 0.5~\mu\Omega$~cm in previous studies of Nair et al. \cite{precursor} and $\sim 1.6~\mu\Omega$~cm in magnetic field of 12~T of Capan et al. \cite{Capanheatcapacity}. Top right panel shows the data in zero field and at 4~T plotted vs. $T^n$ with $n$=1.21, best fit to power-law function for $T_{cf}<T<3.5~K$. This value is very close to previously reported $n$=1.27 \cite{Ronning2014}. Similar power-law dependence with $n$=1.21 is found for fields up to 11~T, with the range of power-law dependence continuing down to $T\approx$ 1~K. At lower temperatures, all $\rho(T)$ curves show clear signs of saturation when $T$ tends to zero, and at the lowest temperatures follow expectations of the Fermi-liquid behavior. This is illustrated in the bottom left panel of Fig.~\ref{fig:resistivity} in which resistivity data are plotted vs. $T^2$, linearizing the dependence with the slope proportional to $A$ coefficient. Slight upward deviations at the lowest temperatures for 8~T and 11~T curves are similar to those observed in CeCoIn$_5$ \cite{Paglione-QCP} and are due to orbital magnetoresistance. For all magnetic fields $H>H_{c2f}$, the $\rho(T^2)$ plots reveal a range of linear dependence at the lowest temperatures. Deviations from $T^2$ dependence at high temperatures mark a characteristic temperature $T_{FL}$, denoted by arrows. Note that $T_{FL}$ has values in the range 0.7 to 1 ~K, with resistivity crossing-over to $T^n$ at slightly higher temperatures. $T_{FL}$, as determined from resistivity measurements, increases slightly with magnetic field, as shown in the phase diagram (bottom right panel of Fig.~\ref{fig:resistivity}). Due to a crossover character of deviations, definitions of $T_{FL}$ have somewhat high errors, so it is difficult to make reliable extrapolation of $T_{FL}(H)$ towards bulk $H_{c2} \approx$ 0.49~T, determined from both thermal conductivity measurements \cite{hybrid,Capanheatcapacity} and heat capacity measurements \cite{Kittaka}. The line in the bottom right panel of Fig.~\ref{fig:resistivity} suggests extrapolation of $T_{FL}$ to finite values in $H=0$ limit, suggesting non-existence of the field-tuned quantum critical point at $H_{c2}$ or even in zero field. However, significant error bars of $T_{FL}$ determination make this extrapolation from field 4~T unreliable towards bulk $H_{c2}$.

To get additional insight into the magnetic field-induced evolution of the normal state in CeIrIn$_5$, we performed thermal conductivity measurements. These measurements are not affected by filamentary superconductivity, and they can be performed in the very proximity of bulk $H_{c2}$. In Fig.~\ref{fig:heat} we show the thermal conductivity of CeIrIn$_5$ measured in three magnetic fields : 0.52~T (slightly above the bulk $H_{c2} = 0.49$~T), 4~T (above $H_{c2f}$), and 10.7~T (far away from superconductivity). As can be seen from 0.52~T curve, the $\kappa/T$ vs $T$ plot does not show any features on entering filamentary zero resistance state below approximately 1.0~K (not shown), leading to artificial violation of Wiedemann-Franz law. As long as filamentary superconductivity is suppressed, thermal conductivity and electrical resistivity perfectly obey Wiedemann-Franz law in the $T \to$~0 limit, with $\kappa /T \sigma=L_0$, where electrical conductivity $\sigma=1/\rho$ and $L_0=\frac{\pi^2}{3}(\frac{k_B}{e}^2$)=2.45$\times$10$^{-8}$W$\Omega$K$^{-2}$ is Sommerfeld value of the Lorenz number, expected for Fermionic systems. The validity of this ratio in $T \to$~0 limit can be seen by direct comparison of the temperature-dependent thermal analog of electrical resistivity, $w\equiv L_0T/\kappa$, and resistivity (right bottom panel in Fig.~\ref{fig:heat}), which meet at $T\to$~0. CeIrIn$_5$ is a good metal, and therefore phonon contribution to thermal conductivity can be completely neglected in the temperature interval of interest, $T<$1~K. In Fig.~\ref{fig:heat} we show this explicitly by subtracting phonon contribution $\kappa_g$ as determined from measurements in disordered samples, from measured $\kappa$ of pure CeIrIn$_5$. The resultant electronic part of thermal resistivity $w_e$ is plotted in bottom right panel of Fig.~\ref{fig:heat} with solid line. Close match is observed between $w$ and $w_e$ below 1~K, validating our conclusion about negligible phonon correction.

\section{Discussion}

\subsection{Temperature-dependent resistivity}

As can be seen from top panels of Fig.~\ref{fig:resistivity}, $\rho(T)$ curves in CeIrIn$_5$ strongly deviate from expectations of the Fermi-liquid functional dependence above $T \sim$ 1~K and are well described by a $\Delta \rho \propto T^n$ dependence with $n$=1.21.  
Observation of very strong $\rho(T)$ dependence at so low temperatures (where phonon scattering becomes negligible but the strength of inelastic scattering, $\Delta \rho _{in}= \rho(1K)-\rho(0)=0.6 $ $ \mu \Omega$cm, is significantly larger than $\rho(0)=0.3$ $\mu \Omega$cm) suggests strong magnetic scattering in the compound, similar to CeCoIn$_5$ \cite{Petrovic-Co,Settai}.  
The power-law non-Fermi-liquid temperature dependent resistivity with similar exponents $n$ are not unusual in the heavy fermion systems close to quantum critical points. For example similar values of the power-law exponents are found in CePd$_2$Si$_2$ ($n \sim 1.2$) \cite{Mathur}.  
The low exponent $n$=1.21 finds explanation within spin fluctuation theory \cite{qcpinside,Hertz} for nearly antiferromagnetic metals in the framework of a spin density wave (SDW) scenario, though significantly lower $\Delta \rho_{in}/\rho_0$ are usually expected in this case \cite{Hertz}. 
In three-dimensional SDW scenario the exponents are usually expected to be close to $n$=1.5 as found in 3D CeIn$_3$ ($n \sim 1.5$) \cite{Mathur}. For the case of two-dimensional magnetic fluctuations the exponents are usually expected to be close to $n$=1 \cite{Rosch}, as found in YbRh$_2$Si$_2$ \cite{Moriya,Gegenwart} and CeCu$_{5.2}$Ag$_{0.8}$ ($n \sim 1$) \cite{Custers}. In CeCoIn$_5$ the $\rho(T)$ in zero field is close to $T$-linear above $T_c$ \cite{Petrovic-Co}, but turns to $n$=1.5 at field-tuned QCP \cite{nonvanishing}. In this compound exponent $n$ reveals significant anisotropy with $n$=1 in $\rho_c(T)$ \cite{Malinowski,Heuser}.

In CeIrIn$_5$ the same power law behavior of resistivity up to 3.5~K is found for all magnetic fields. This observation is suggestive, that despite magnetic character of scattering, dominant scattering is not tuned by magnetic field. This is in stark contrast to CeCoIn$_5$ which shows different power law resistivity for low and high magnetic fields in the normal state \cite{Paglione-QCP,nonvanishing}.

\begin{figure*}[t] 
\centerline{ 
\scalebox{0.40}{ 
\includegraphics{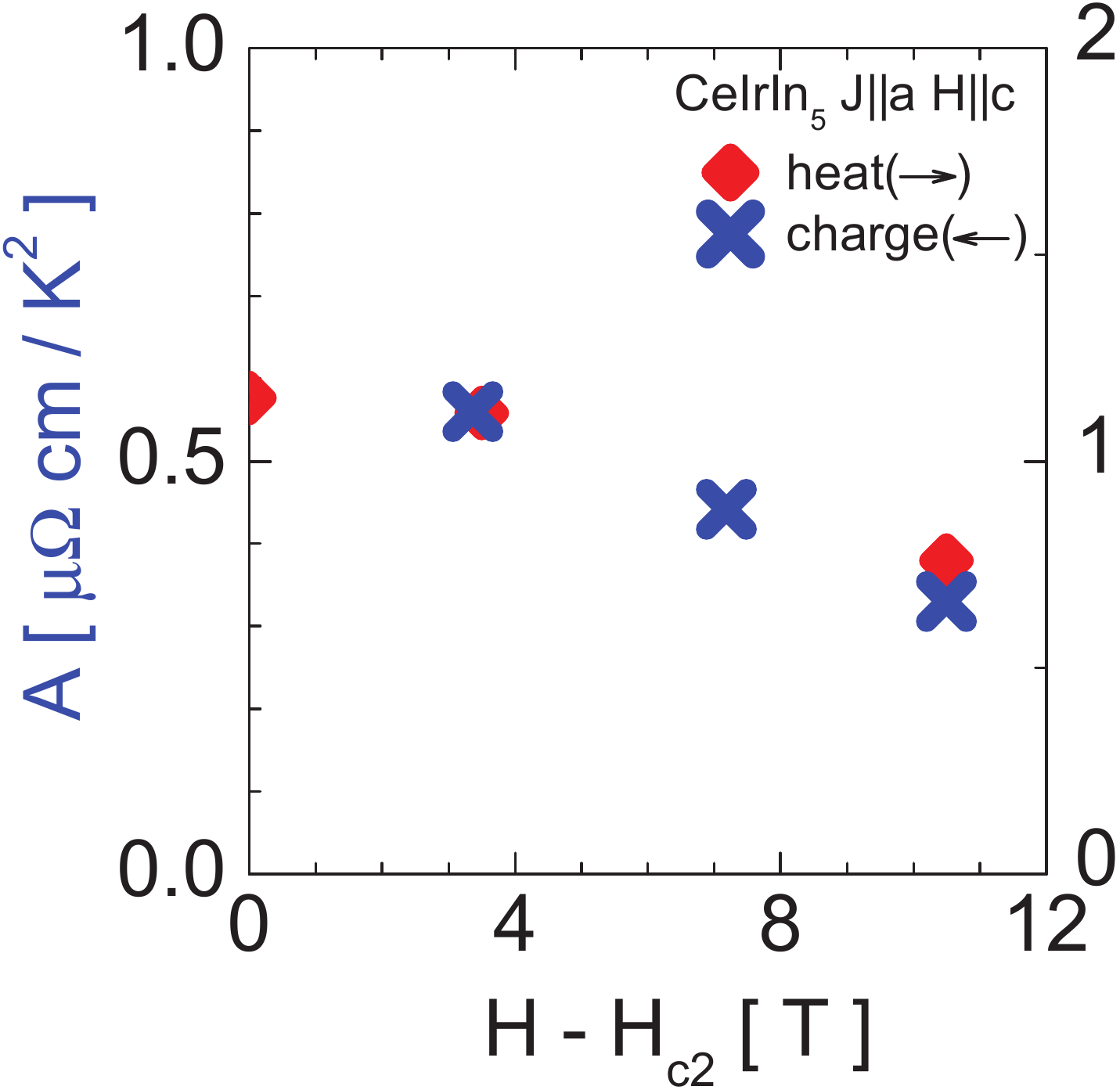}} 
\scalebox{0.40}{ 
\includegraphics{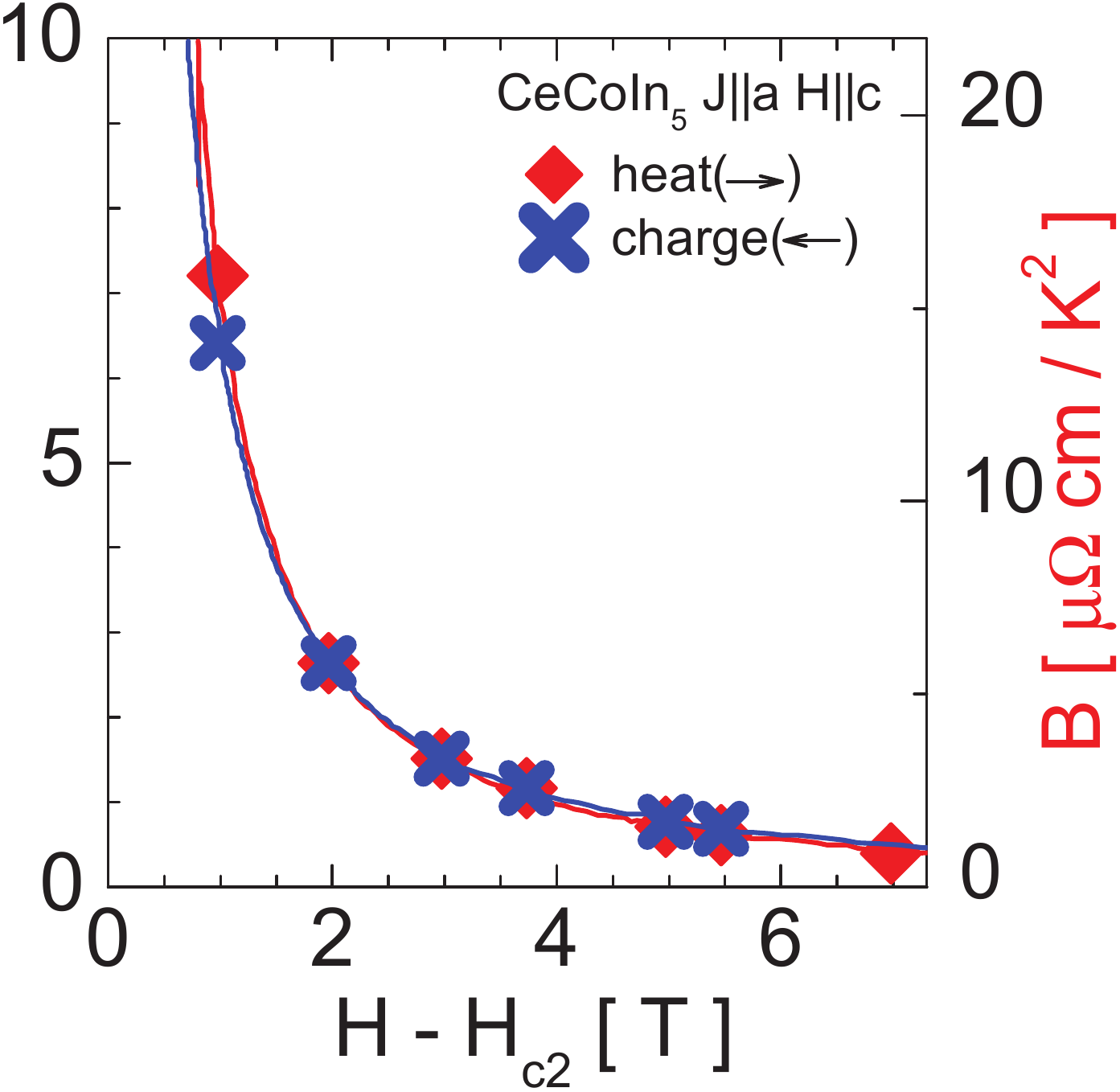}}} 
\centerline{ 
\scalebox{0.40}{ 
\includegraphics{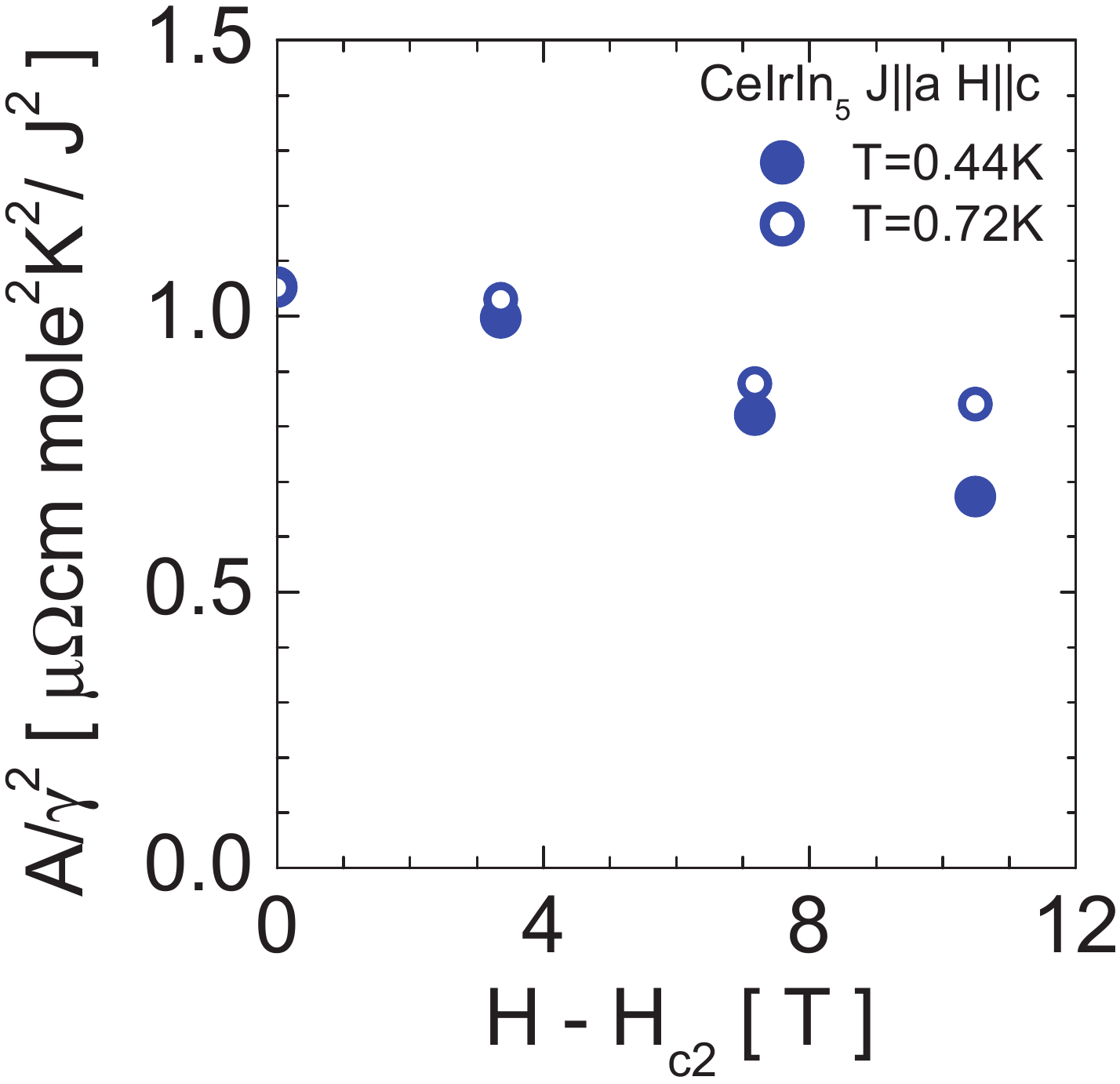}} 
\hspace*{0.55cm} 
\scalebox{0.40}{ 
\includegraphics{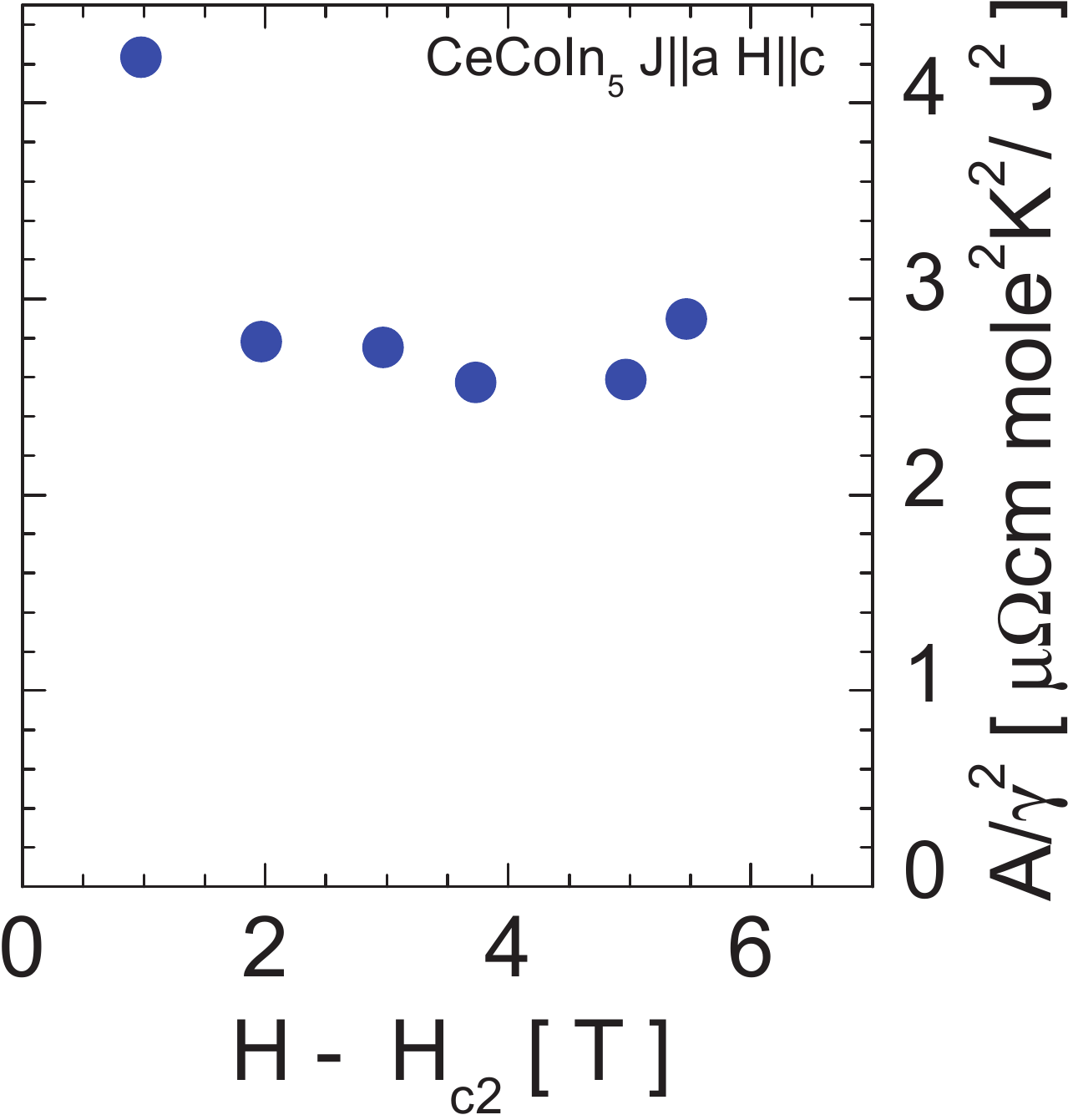}}} 
\caption{\label{fig:AvsB} 
Magnetic field dependence of the $T^2$ coefficients of the electrical resistivity, $A$ (blue crosses, left axes), and of the thermal resistivity, $B$ (red diamonds, right axes) for CeIrIn$_5$ (top left panel) and CeCoIn$_5$ (top right panel). The data are plotted versus $H-H_{c2}$, distance to a quantum critical point in CeCoIn$_5$. Note ten-fold difference of the $A/B$ scales for the two cases. The field dependence $B(H)$ in CeIrIn$_5$ suggests no divergence of the effective mass towards bulk $H_{c2}$. Significantly smaller values of $A$ and $B$ coefficients compared to those in CeCoIn$_5$ suggest CeIrIn$_5$ is significantly further away from quantum critical point in the phase diagram. Decrease of the $A$ and $B$ coefficients with magnetic field suggests moving away from QCP. Bottom panels show Kadowaki-Woods ratios, $A/ \gamma ^2$, calculated based on the measured $A$ values and the literature values of the specific heat \cite{Capanheatcapacity}. Note that heat capacity data show notable temperature dependence at 11~T, as reflected in the higher ratio value at 0.72~K.  } 
\end{figure*} 
  
\subsection{Field evolution of the effective mass} 
  
Temperature-dependent part of electronic thermal resistivity obeys the same functional dependence as electrical resistivity, however, because of the different Fermi surface integrals of heat and charge scatterings, the $T^2$ coefficients $A$ for resistivity and $B$ for thermal resistivity are not the same. Typically, for Fermi surfaces without sharp corners, the two coefficient are different by a factor of about two \cite{Paglione-Rh}. In the left panel of Fig.~\ref{fig:AvsB} we plot magnetic field dependence of $A$ and $B$ coefficients of the $T^2$ terms in the in-plane electrical (left axis) and thermal (right axis) resistivity of CeIrIn$_5$. The $A$ and $B$ coefficients in CeIrIn$_5$ are indeed close to obeying $B=2A$ relation, similar to CeCoIn$_5$ and most other cases \cite{WFScience,B2A1,B2A2,B2A3}. The $A$ and $B$ coefficients in CeIrIn$_5$ closely follow each other as functions of magnetic fields, with $B$ decreasing from approximately 1.2 to 0.8 on field increase from 0.52~T to 11~T. For reference in the right panel we show similar data for CeCoIn$_5$ \cite{nonvanishing}. Note ten-fold difference in the scales for the two compounds. In sharp contrast to CeCoIn$_5$, $A$ and $B$ coefficients in CeIrIn$_5$ do not reveal any tendency for divergence as functions of magnetic fields. The nearly ten-fold difference in $A$ and $B$ coefficients between CeCoIn$_5$ and CeIrIn$_5$ suggest notable difference in the effective masses, reflecting close proximity to quantum critical point in case of CeCoIn$_5$ and relatively large distance from it for CeIrIn$_5$.

This very weak field dependence of $A$ and $B$ coefficients in CeIrIn$_5$ is consistent with specific heat \cite{Capanheatcapacity,Stewart}, finding weak field dependence of the electronic specific heat coefficient $\gamma_{0}$ , and the dHvA experiments \cite{Shishido-Haga} and the cyclotron effective mass, $m^{*}$, (for field between 6 to 17 T) at low temperature \cite{Capanheatcapacity,Shishido-Haga}. In CeCoIn$_5$ $A$ coefficient changes from 7.5 to 0.5 $~\mu\Omega$~cm~K$^{-2}$ for fields $\sim 6$ to 16 T) \cite{Paglione-QCP}, and significant field dependence is found for effective masses of carriers, particularly strong for $\beta$ and $\alpha$ sheets of the Fermi surface \cite{Settai}.

In our study of field-tuned quantum criticality in CeCoIn$_5$ on approaching from both inside the superconducting dome from below $H_{c2}$ and from outside the superconducting dome from above $H_{c2}$ \cite{qcpinside} we noticed that the most heavy electrons are strongly bound in the superconducting condensate. In this case the big difference in the strength of electron-electron renormalization between CeCoIn$_5$ and CeIrIn$_5$, can be directly linked to five times difference in the value of the superconducting $T_c$. CeCoIn$_5$ and CeIrIn$_5$ have identical Fermi surfaces \cite{bandstructureFS}, yet interestingly demonstrate very different electron renormalization and $T_c$.

Another interesting observation is that the transition at $H_{c2}$ in CeIrIn$_5$ is second order \cite{Kittaka}, while in CeCoIn$_5$ it is first order. First order character of the transition is taken as arising from paramagnetic limiting of the upper critical field. We can speculate that orbital limiting of $H_{c2}$ observed in CeIrIn$_5$ in configuration $H \parallel c$ can be responsible for the lack of coincident quantum critical point.

Summarizing, we neither find field-tuned quantum critical point at bulk $H_{c2}$ in CeIrIn$_5$, nor see indications of its existence in all range of magnetic fields studied (up to 11~T). This is in sharp contrast to CeCoIn$_5$, suggesting that CeIrIn$_5$ is significantly further away from QCP. This finds support in a ten-fold difference of the Fermi liquid $T^2$ coefficients between the two compounds.

\subsection{Kadowaki-Woods ratio} 
 
It is of interest to note that very big difference in $A$ and $B$ coefficients between CeCoIn$_5$ and CeIrIn$_5$ is not expected based on previous specific heat studies \cite{Stewart,Capanheatcapacity,Kittaka}.  
In zero field both materials have similar $\gamma=C/T$ values above $T_c$, somewhat higher in CeInIn$_5$. 
Big difference in $A$ coefficients suggests that the Kadowaki-Woods ratio, defined as $r_{KW}=A/\gamma^2$, is anomalously low in CeIrIn$_5$.  
Indeed, as shown in Fig.~\ref{fig:AvsB} near $H_{c2}$ in CeIrIn$_5$ we find $r_{KW}\sim 0.1 a_{0}$ ($a_0$=10$~\mu\Omega$~cm~mol$^{2}$~K$^{2}$/~J$^{2}$), and the ratio does not show significant field dependence.  
In CeCoIn$_5$, for instance, this ratio is about 0.52 $a_0$ at $H_{c2} \sim 6 ~T$ that falls on the universal line for some other heavy fermions \cite{Paglione-QCP}.

The Kadowaki-Woods ratio cancels out mass renormalization in electrical resistivity and heat capacity measurements and leads to scaling of $A$ and $\gamma^2$ proportional to level degeneracy of $f$-electrons \cite{ColemanKW}. Ce ions in CeCoIn$_5$ and CeIrIn$_5$ are in the same crystal field environment, so it is natural to take the same value $n$=2. While the $r_{KW}$ for CeCoIn$_5$ has a value typical for heavy fermion materials, it is anomalously low in CeIrIn$_5$.

Anomalously low value of Kadowaki-Woods ratio in CeIrIn$_5$ suggests that not the same groups of carriers are determining the electrical resistivity and the heat capacity of the samples. Two scenarios can be at play. In the first one, conduction electrons are strongly disconnected from $f$-electrons due to localized character of the latter. In this case similar values of heat capacity in CeCoIn$_5$ and CeIrIn$_5$ reflect nearly identical $f$-electron contribution, but their significantly more localized character in CeIrIn$_5$. In the second scenario, we need to assume significant difference of properties of different groups of carriers on the multi-band Fermi surface of the materials. This scenario would require the presence of light and mobile carriers dominating in the electrical transport and heavy carriers dominating the heat capacity.

One possible way to distinguish these two scenarios is to study magnetoresistance. In the latter scenario the magnetoresistance of CeIrIn$_5$ should be significantly larger than in CeCoIn$_5$. Comparison of the resistivity data for CeIrIn$_5$ in magnetic field of 11~T in Fig.~\ref{fig:resistivity} bottom left panel with similar data in CeCoIn$_5$ \cite{Paglione-QCP} we find that orbital magnetoresistance at the lowest temperatures is stronger in CeCoIn$_5$ which is at odds with this scenario.

\section{Conclusions} 


In conclusion, temperature-dependent electrical resistivity and thermal conductivity measurements in CeIrIn$_5$ as functions of magnetic fields applied along $c$-axis revealed the validity of the Wiedemann-Franz law in $T \to 0$ limit as long as resistivity measurements are not affected by filamentary superconductivity.  
At temperatures above 1~K the temperature-dependent resistivity is well represented by a power-law function $\Delta \rho \propto T^n$, with nearly field-independent $n$=1.21.  
At the lowest temperatures the temperature dependences of both electrical resistivity, $\rho(T)$ and of thermal resistivity, $\rho (T)$, follow expectations of the Landau Fermi-liquid theory with $B \approx 2A$, as expected for Fermi surfaces without sharp features in the angular Fermi velocity distributions. 
The coefficient $B$ does not show significant field dependence even on approaching $H_{c2}$=0.4~T of bulk superconducting state. Weak response to magnetic field is in stark contrast with the behavior found in the closely related CeCoIn$_5$, in which field-tuned quantum critical point coincides with $H_{c2}$. The value of the electron-electron mass enhancement, as judged by $A$ and $B$ coefficients, is about one order of magnitude reduced in CeIrIn$_5$ as compared to CeCoIn$_5$, suggesting that the material is significantly further away from magnetic quantum critical point. Suppressed Kadowaki-Woods ratio in CeIrIn$_5$ compared to CeCoIn$_5$ suggests notably more localized nature of $f$-electrons in the compound.


\subsection{Acknowledgement} 
We thank J. Corbin for his assistance with the experiment and R. Flint for useful discussions. L.T. acknowledges support from the Canadian Institute for Advanced Research and funding from NSERC, FRQNT, CFI, and the Canada Research Chairs Program. Part of the work was carried out at the Brookhaven National Laboratory, which is operated for the U.S. Department of Energy by Brookhaven Science Associates (No. DEAC02-98CH10886). H.Sh. would like to acknowledges Iran National Science Foundation (INSF) for supporting this project.

  
\end{document}